\documentclass[letterpaper]{article} 
\usepackage{aaai2026}  
\usepackage{times}  
\usepackage{helvet}  
\usepackage{courier}  
\usepackage[hyphens]{url}  
\usepackage{graphicx} 
\usepackage{natbib}  
\usepackage{caption} 
\usepackage{booktabs}
\usepackage{amssymb}
\usepackage{amsmath}
\usepackage{multirow}
\usepackage{xcolor}
\usepackage{algorithm}
\usepackage{algorithmic}

\usepackage{newfloat}
\usepackage{listings}
\DeclareCaptionStyle{ruled}{labelfont=normalfont,labelsep=colon,strut=off} 
\floatstyle{ruled}
\newfloat{listing}{tb}{lst}{}
\floatname{listing}{Listing}
\title{Video Echoed in Music: Semantic, Temporal, and Rhythmic Alignment for Video-to-Music Generation}
\author{
    Xinyi Tong\textsuperscript{\rm 1,2,3}\equalcontrib,
    Yiran Zhu\textsuperscript{\rm 3}\equalcontrib,
    Jishang Chen\textsuperscript{\rm 1,2,3},
    Chunru Zhan\textsuperscript{\rm 3},
    Tianle Wang\textsuperscript{\rm 1,2},
    Sirui Zhang\textsuperscript{\rm 1,2},\\
    Nian Liu\textsuperscript{\rm 2},
    Tiezheng Ge\textsuperscript{\rm 3},
    Duo Xu\textsuperscript{\rm 2},
    Xin Jin\textsuperscript{\rm 2},
    Feng Yu\textsuperscript{\rm 1},
    Song-Chun Zhu\textsuperscript{\rm 2,4}\thanks{Corresponding authors.}
}
\affiliations{
    \textsuperscript{\rm 1}Central Conservatory of Music, Beijing, China\\
    \textsuperscript{\rm 2}Beijing Institute for General Artificial Intelligence, Beijing, China\\
    \textsuperscript{\rm 3}Alibaba Group, Beijing, China\\
    \textsuperscript{\rm 4}Peking University, Beijing, China\\
    tongxinyi@mail.ccom.edu.cn, jinxinbesti@foxmail.com, s.c.zhu@pku.edu.cn
}

\begin{document}

\maketitle

\begin{abstract}
Video-to-Music generation seeks to generate musically appropriate background music that enhances audiovisual immersion for videos. However, current approaches suffer from two critical limitations: 1) incomplete representation of video details, leading to weak alignment, and 2) inadequate temporal and rhythmic correspondence, particularly in achieving precise beat synchronization. To address the challenges, we propose \textbf{V}ideo \textbf{E}choed in \textbf{M}usic (VeM), a latent music diffusion that generates high-quality soundtracks with semantic, temporal, and rhythmic alignment for input videos. To capture video details comprehensively, VeM employs a hierarchical video parsing that acts as a music conductor, orchestrating multi-level information across modalities. Modality-specific encoders, coupled with a storyboard-guided cross-attention mechanism (SG-CAtt), integrate semantic cues while maintaining temporal coherence through position and duration encoding. For rhythmic precision, the frame-level transition-beat aligner and adapter (TB-As) dynamically synchronize visual scene transitions with music beats. We further contribute a novel video-music paired dataset sourced from e-commerce advertisements and video-sharing platforms, which imposes stricter transition-beat synchronization requirements. Meanwhile, we introduce novel metrics tailored to the task. Experimental results demonstrate superiority, particularly in semantic relevance and rhythmic precision.
\end{abstract}

\section{Introduction}

Music, akin to video, evokes sensory perception and emotional responses. This intrinsic relationship underscores the integration to enhance the audiovisual experience. However, manual music composition is time-consuming and costly, while relying on pre-existing music raises issues of copyright and congruence. Thus, Video-to-Music(V2M) generation presents a promising solution with broad applications in film, advertising, gaming, and short-form video production.

\begin{figure}[ht]
\centering
\includegraphics[width=\linewidth]{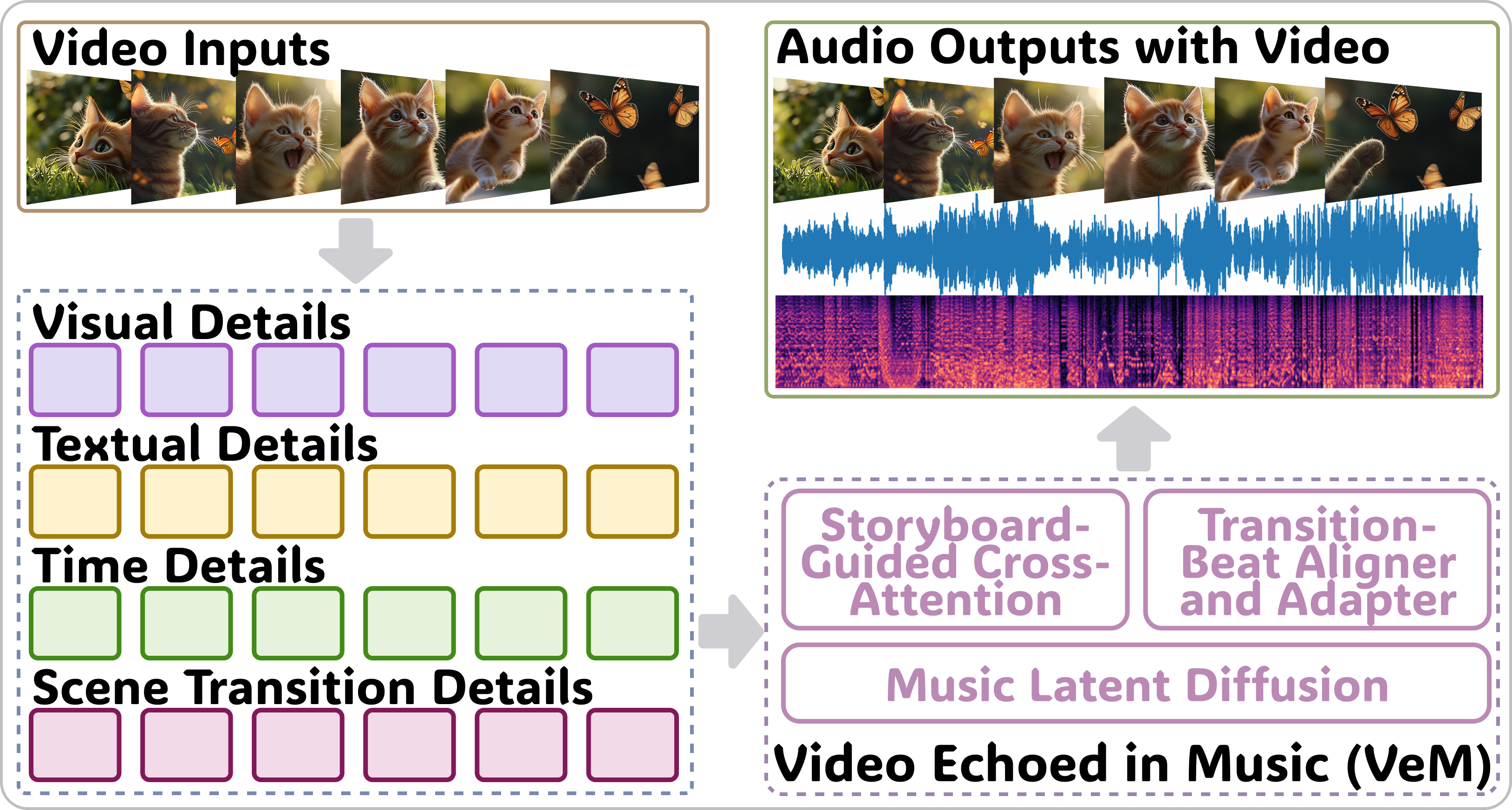}
\caption{\textbf{Task overview.} The proposed latent music diffusion, VeM, achieves semantic, temporal, and rhythmic alignment during video-to-music generation by integrating multimodal details derived from videos as conditions.}
\label{fig:fig1}
\end{figure}

The V2M task aims to generate background music that exhibits semantic, temporal, and rhythmic alignment with the given video. This involves three critical aspects: 1) \textbf{High fidelity} ensures that music is indistinguishable from human-composed pieces, serving as a fundamental benchmark for music generation. 2) \textbf{Semantic alignment}, whereby music accurately reflects thematic, emotional, and narrative elements in videos. 3)\textbf{Temporal synchronization} emphasizes alignment with temporal dynamics by jointly integrating semantic and temporal cues during generation. \textbf{Rhythmic consistency}, as a distinctive dimension of temporal alignment, accentuates junctures by synchronizing video transitions with music beats, ensuring transition-beat matching.

Recent research has advanced in these areas. 
1) For music quality, some focus on symbolic representations to meet human-composed standards, but audio synthesis with engines restricts timbral diversity \cite{di2021video, zhuo2023video,xie2025filmcomposer}. More efforts \cite{agostinelli2023musiclm,copet2023simple} shift towards waveform directly, facilitating superior auditory feedback, which we have also adopted. 2) Existing semantic alignment methods fall into two broad categories. The first employs rule-based or learnable visual features to guide generation \cite{yu2023long,li2024diff,xie2025filmcomposer, tian2025xmusic}. However, the features provide coarse video understanding, potentially imposing insufficient constraints. Although MuVi \cite{li2024muvivideotomusicgenerationsemantic} and VidMusician \cite{li2024vidmusician} advance with visual adapters, they neglect global semantic invariance over time. The second category leverages visual-language models to extract textual descriptions \cite{tian2024vidmuse,tong2024video, wang2024multimodal,zhou2025harmonyset}, reducing the task to text-to-music and largely bypassing visual features. The inherent limitations of text hinder temporal details, leading to poor synchronization. 3) For temporal synchronization, recent methods employ local semantics to involve temporal variations through video clips \cite{li2024diff,zuo2025gvmgen} or textual timestamps \cite{zhang2025sonique,zhou2025harmonyset}, but typically overlook fine-grained temporal details. More works focus on rhythmic consistency by aligning partial visual dynamics with musical rhythms, including optical flow \cite{di2021video,KANG2024123640}, visual embedding variation \cite{zhuo2023video,lin2024vmas,li2024vidmusician,xie2025filmcomposer}, and human-centric motion \cite{zhu2022dance,li2024dance,you2024momu}. These specific dynamics fail to explicitly capture the rhythmic cues. 

The most pertinent research, Video-to-Audio, generating sound effects from videos, also emphasizes temporal consistency \cite{ruan2023mm,liu2024audioldm,luo2024diff,xing2024seeing,wang2024frieren,rong2025audiogenie}. However, applying the strategy directly to music presents challenges. Sound effects align with discrete visual events, whereas music exhibits intrinsic rhythmic periodicity with recurring beats, requiring longer alignment spans and smoother transitions. Crucially, salient video transitions typically coincide with music beats; arbitrary deviations can disrupt the rhythmic flow and lead to discordance.

In this paper, we propose \textbf{V}ideo \textbf{e}choed in \textbf{M}usic (VeM), a diffusion-based framework to achieve semantic, temporal, and rhythmic alignment for V2M generation. We provide a hierarchical video parsing, serving as a music conductor, which comprehensively orchestrates multilevel details,  shown in Fig. \ref{fig:fig1}. Semantic and temporal cues are integrated by a storyboard-guided cross-attention mechanism (SG-CAtt). Rhythmic precision is maintained by frame-level transition-beat aligner and adapter (TB-As), synchronizing video transitions with music beats. Meanwhile, we construct TB-Match, a video-music paired dataset collected from e-commerce advertisements and video-sharing platforms, enforcing stricter synchronization for transitions and beats. We introduce novel evaluation metrics tailored to the task. The experimental results demonstrate superiority in both semantic-temporal relevance and rhythmic precision. The main contributions are claimed as follows:

\begin{itemize}
    \item A novel perspective that utilizes hierarchical video parsing as a music conductor to orchestrate comprehensive multimodal constraints for video-to-music generation.
    \item A diffusion-based framework that explicitly integrates multimodal constraints into soundtracks to achieve semantic, temporal, and rhythmic alignment.
    \item A video-music dataset annotated with fine-grained parsing and evaluation metrics tailored to the task. Both subjective and objective results show the superiority.
    
\end{itemize}

\section{Related Works} 

\begin{figure*}[ht]
\centering
\includegraphics[width=\linewidth]{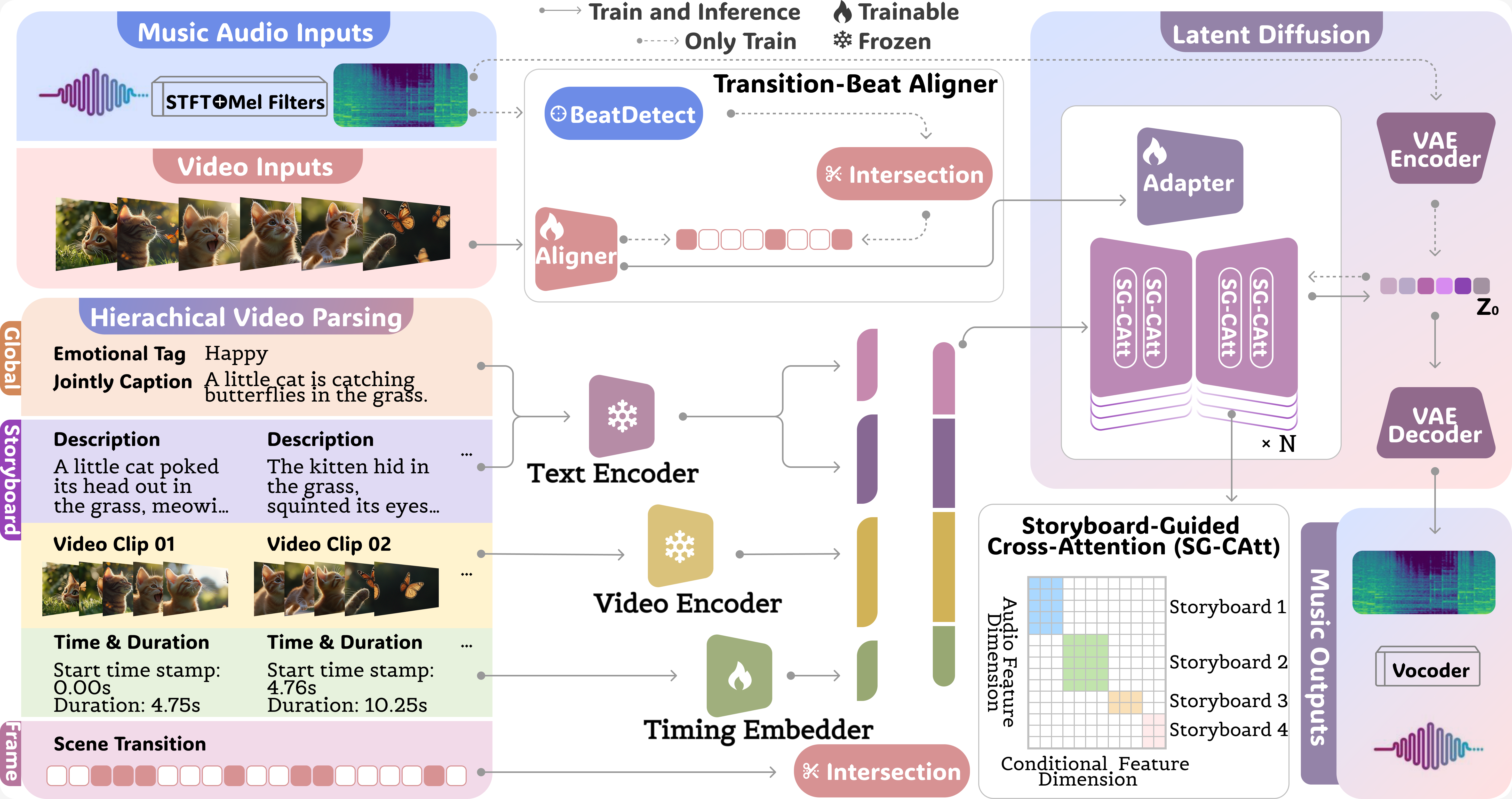}
\caption{\textbf{Illustration of the proposed method.} The \textbf{hierarchical video parsing} provides a comprehensive analysis across three levels. Cross-modal features are captured by \textbf{modality-specific encoders}, facilitating the semantic and temporal alignment by integrating global and storyboard details into the generative latent via \textbf{storyboard-guided cross-attention}. The frame-level \textbf{transition-beat aligner and adapter} ensure precise rhythmic synchronization by coupling video scene transitions with detected music beats and adapting to the music latent.}
\label{fig:fig2}
\end{figure*}

\subsection{Diffusion-Based Conditional Music Generation}

Recent advances in diffusion models have demonstrated potential for conditional music generation. Riffusion \cite{forsgren6riffusion}, Noise2Music \cite{huang2023noise2music}, and Mo\^{u}sai \cite{schneider2023mo} have pioneered open-domain text-to-music generation by diffusion models. AudioLDM2 \cite{liu2024audioldm} facilitates holistic audio generation, including music, through self-supervised pretraining. DITTO \cite{novack2024ditto} leverages distilled diffusion inference-time T-optimization for enhanced generation. Mustango \cite{melechovsky2024mustango} and Music ControlNet \cite{wu2024music} apply various time-varying musical constraints (e.g., chords, rhythms), while MusicMagus \cite{zhang2024MusicMagus} and SteerMusic \cite{niu2025steermusic} explore zero-shot music editing via diffusion. These developments underscore the effectiveness of diffusion models for conditional music generation. Building upon the foundations, we present VeM that extends latent diffusion to video-to-music while retaining the controllability benefits established in conditional music generation.

\subsection{Video-to-Music Generation}
Current approaches for video-to-music alignment employ diverse strategies. The first method, CMT \cite{di2021video} and subsequent approaches \cite{app12105050,zhuo2023video,yu2023long,KANG2024123640,qi2024harmonizing} project disentangled visual features (RGB, saliency, motion) onto musical attributes (melody, chord, rhythm), failing to capture visual semantics. Large Language Model-based techniques \cite{liu2023m,xu2024mozart,tong2024video,wang2024multimodal,zhou2025harmonyset} leverage textual representations. Specifically, M$^2$UGen \cite{liu2023m} focuses on textual music understanding, while SONIQUE \cite{zhang2025sonique} extracts musical tags from unpaired data. AudioX \cite{tian2025audiox} combines visual, textual, and audio features to a multimodal condition. However, textual abstraction inherently loses fine-grained temporal dynamics. Motion-centric methods, such as V2Meow \cite{su2024v2meow}, FilmComposer \cite{xie2025filmcomposer}, and VMAS \cite{lin2024vmas}, achieve movement alignment but neglect broader domains. VidMuse \cite{tian2024vidmuse} involves long-short-term temporal dependencies, but suffers from limited generative capacity. Diff-BGM \cite{li2024diff} addresses clip-level alignment, but only partially adapts to semantic shifts. Recent approaches, MuVi \cite{li2024muvivideotomusicgenerationsemantic}, VidMusician \cite{li2024vidmusician}, and GVMGen \cite{zuo2025gvmgen}, improve local semantic correspondence that involves temporal dynamics but lack explicit temporal position and duration encoding, preventing precise frame-level synchronization. Therefore, substantial opportunities remain for advancing semantic, temporal, and rhythmic alignment in video-to-music generation.

\section{Method}

This section introduces the proposed VeM, a latent music diffusion to achieve semantic, temporal, and rhythmic alignment for videos. The pipeline is illustrated in Fig. \ref{fig:fig2} \footnote{Demos and Code are available at \url{https://vem-paper.github.io/VeM-page/}}. Hierarchical video parsing acts as a music conductor, providing comprehensive multimodal video details that are represented by modality-specific encoders. Semantic and temporal cues are integrated via SG-CAtt. Fine-grained rhythmic precision is ensured through frame-level TB-As.

\subsection{Preliminary}

\textbf{Music Audio Representation}. 
For a music waveform $x\in\mathbb{R}^{L_s}$, where $L_s$ denotes the number of audio samples, we adopt the log Mel-spectrogram $X \in\mathbb{R}^{W \times B}$ as the training target, derived via the Short-Time Fourier Transform (STFT) and Mel-filters, due to its perceptual relevance and dimensionality reduction. $W$ and $B$ represent the time windows and Mel-frequency bins, respectively. A trained variational autoencoder (VAE) encodes $X$ into a latent representation $z$. We subsequently train a latent diffusion to generate $z$ by iteratively denoising from Gaussian noise $\epsilon$. Finally, the predicted latent $z$ is reconstructed to the Mel-spectrogram by the VAE decoder, followed by waveform synthesis via the vocoder \cite{kong2020hifi}.

\textbf{Latent Music Diffusion} 
The Latent Diffusion Model (LDM) \cite{rombach2022high} comprises the diffusion phase and the denoising phase. The forward diffusion phase is a $T$-step Markov process that corrupts the input by iteratively adding noise to a standard isotropic Gaussian distribution. Given latent $z_{t-1}$ at step $t-1$, the distribution of $z_t$ at step $t \in 2, ... , T$ is defined as:
\begin{equation}
    \label{eq1}
    q(z_t|z_{t-1})=\mathcal{N}(z_t;\sqrt{1-\beta}z_{t-1}, \beta_t\mathbf{I})
\end{equation}
where the noise schedule hyperparameter, $\beta_t\in[0, 1]$, regulates the rate at which noise is applied to the data. By recursively substituting $q(z_t|z_{t-1})$, the formulation is derived:
\begin{equation}
    \label{eq2}
    q(z_t|z_0)=\mathcal{N}(z_t;\sqrt{\overline{\alpha}_t}z_0, (1-\overline{\alpha}_t)\mathbf{\epsilon}
\end{equation}
where $\alpha_t$ parameterizes $1-\beta_t$ and $\overline{\alpha}_t := \prod^t_{s=1}\alpha_s$ represents the cumulative noise level at timestep $t$. $z_T\sim\mathcal{N}(0,\mathbf{I})$ indicates the final state at step $T$ follows a standard isotropic Gaussian distribution. $\mathbf{\epsilon} \sim \mathcal{N}(0,\mathbf{I})$ denotes noise addition. During the reverse process, we implement a Transformer-UNet (T-UNet) architecture, which is crafted to optimize the noise estimation objective:
\begin{equation}
    \label{eq3}
    \mathcal{L}=\mathbb{E}_{z_0,\epsilon\sim\mathcal{N}(0, 1), t, c}[\|\epsilon-\epsilon_{\theta}(z_t,t,c)\|^2]
\end{equation}
The process iteratively generates the prior \(z_0\) according to: 
 \begin{equation}
    \label{eq4}
     p_\theta(z_{0:T}|c)=p(z_T)\prod^T_{s=t}p_\theta(z_{t-1}|z_t,c)
 \end{equation}
 \begin{equation}
    \label{eq5}
    p_\theta(z_{t-1}|z_t, c)=\mathcal{N}(z_{t-1};\mu_\theta(z_t,n,c),\sigma_t^2\mathbf{I})
\end{equation}
where $\epsilon_{\theta}(z_t,t,c)$ is the predicted noise, $\mu_\theta$ and $\sigma_t$ denote parameterized mean and variance, and $c$ stands for conditions provided to the model. During the training phase, T-UNet is optimized to learn a backward transition from the prior distribution $\mathcal{N}(0, \mathbf{I})$ to the target $z$, conditioned on the input $c$. In this paper, we structure hierarchical video representation captured by modality-specific encoders as condition signals.

\subsection{Hierarchical Video parsing — Music Conductor}

For comprehensive video analysis, five key elements are supposed to be determined: 1) the overarching theme, atmosphere, and emotional impact; 2) smooth video segmentation into coherent video shots; 3) narrative and visual compositions within each shot; 4) temporal boundaries and duration of each shot; 5) precise timing of frame-level visual changes. The five details are collectively derived from hierarchical video parsing, as depicted in Fig. \ref{fig:fig2}, where segmented shots are conceptualized as storyboards and frame-level changes as scene transitions. 

Hierarchical parsing operates on three levels: global, storyboard, and frame. At the global level, video captions from a video understanding model and emotional tags from a music classification model address Key 1. The storyboard level employs a video segmentation model to extract local visual features, descriptions, start timestamps, and durations, corresponding to Keys 2–4. At the frame level, a scene transition detector ensures precise transitions, enabling fine-grained rhythmic synchronization for Key 5. Details of the aforementioned models are provided in Appendix A. Since video parsing is independent of the training process, we perform it as a preprocessing annotation step, with manual correction and cleaning.

\subsection{Modality-Specific Video Representation}
To fully leverage the rich parsing details of the video, we employ modality-specific encoders for representation. Textual information is encoded using CLAP \cite{wu2023largescale}, a pre-trained text-audio contrastive model. Visual content is processed by MAViL \cite{NEURIPS2023mavil}, which projects videos into a shared video-audio latent space. This strategy ensures consistency between textual and visual embeddings in the audio domain, containing the features of the global video caption $f^{C}_{t}$ and the emotional tag $f^{T}_{t}$, the storyboard-level description $f^{story_i}_{t}$ for the $i$-th storyboard and the corresponding visual features $f^{story_i}_{v}$. Temporal details, including the storyboard start time $f^{story_i}_{s}$ and duration $f^{story_i}_{d}$, are encoded by a learnable continuous-time MLP operating on seconds. Frame-level scene transitions are represented by a binary timestamp indicator $f^{frame-v_i}_{b}$, indicating the presence or absence of a transition for each frame.

\subsection{Storyboard-Guided Cross-Attention}

While cross-attention mechanisms are effective for aligning condition signals with generative representations across modalities \cite{ruan2023mm,tian2025audiox}, existing implementations exhibit critical limitations in temporal modeling. For example, the segment-aware approach \cite{li2024diff} involves local temporal cues, but suffers from rigid segment divisions that neglect natural semantic boundaries. Thus, we propose storyboard-guided cross-attention (SG-CAtt) that explicitly preserves semantic alignment and simultaneously ensures temporal synchronization.

To incorporate global information $f^{C}_{t}$ and $f^{T}_{t}$ into each individual storyboard $i$, we concatenate global features with storyboard-specific features:
\begin{equation}
    f_{att}^i=\left\{f^{C}_{t}\|f^{T}_{t}\|f^{story_i}_{t}\|f^{story_i}_{v}\|f^{story_i}_{s}\|f^{story_i}_{d}\right\}
    \label{eq6}
\end{equation}
For a video with $N$ number of storyboards, the conditional feature is $F_{att}=\left\{f_{att}^1, f_{att}^2, ..., f_{att}^N\right\}$ and serves as the Value and Key within cross-attention. The Query is provided by the latent representation $z_t$ of the diffusion model \cite{vaswani2017attention}. The temporal boundaries are defined by the start time $s^{i}$ and duration $d^{i}$ of the storyboard. To constrain the fusion between the condition and the latent operated solely within relevant storyboards, we introduce a storyboard mask that restricts attention to the interval $[s^{i}, s^{i}+d^{i}]$: 
\begin{equation}
    sMask_{x, y} =\left\{ 
    \begin{array}{ll}
        1, & s^i \leq x, y < s^{i}+d^{i} \\
        0, & else
    \end{array}
    \right\}
\end{equation}
where $x$ and $y$ represent the temporal indices of music latent and conditional features, respectively. As shown in Fig. \ref{fig:fig2}, the mask delineates rectangular regions due to the varying sequence lengths of each storyboard. The SG-CAtt is defined as:
\begin{equation}
\scalebox{0.94}{$
 Attention(Q, K, V) = softmax(sMask \odot \frac{QK^T}{\sqrt{d_{key}}}) \cdot V
 $}
\end{equation}
where $\odot$ denotes element-wise multiplication. Within the T-UNet architecture, the self-attention layers in the final transformer blocks at each level are replaced by the SG-CAtts. To enforce consistent guidance across T-UNet levels, we apply uniform up-sampling and down-sampling ratios, adjusting feature dimensions of the conditional mask. The SG-CAtt technique facilitates semantic alignment and temporal synchronization at the storyboard level. By concatenating global features, semantic consistency is preserved among all storyboards, while masked cross-attention targets local temporal synchronization within individual storyboard boundaries.

\begin{table*}[ht]
\centering
\begin{tabular}{l@{\hspace{20pt}}cc@{\hspace{20pt}}ccccccccc}
    \toprule 
            & Au. & Vd.
            & IS$\uparrow$
            & FAD$\downarrow$  & KLD$\downarrow$ & CLAP$\uparrow$  & LB$\uparrow$ & tw-CLAP$\uparrow$  & tw-LB$\uparrow$ & B$_{IoU}$$\uparrow$ & TB$_{IoU}$$\uparrow$ \\
    \midrule
GroundTruth  &        &        &   -    &   -   &  -    
& 0.247      & 0.928   & 0.252     & 0.932      &  1.000   & 0.559 \\
CMT & $\times$ & $\checkmark$  & 1.131  & 7.151 & 5.540 
& 0.109      & 0.728   & 0.113     & 0.775      &  0.254   & 0.213 \\
Diff-BGM & $\times$ & $\checkmark$   & 1.173   & 6.940    & 4.870   
& 0.112      & 0.781   & 0.109     & 0.792     &  0.227   &  0.261 \\
M$^2$UGen & $\checkmark$ &  $\times$ &1.211    & 5.902    & 3.350  
& 0.158      & 0.892   & 0.163     & 0.893     & 0.307    & 0.331 \\
VidMuse  & $\checkmark$ & $\checkmark$ & 1.206     & 7.437    & 4.210  
&  0.102     & 0.704   &  0.103    & 0.718     & 0.335    & 0.352 \\
GVMGen  & $\checkmark$ & $\checkmark$ & 1.227  & 6.137 & 3.210  
&  0.212     & 0.899   &  0.219    & 0.917     & 0.465   & 0.357 \\
Ours  & $\checkmark$  & $\checkmark$  & \textbf{1.263}  & \textbf{4.043}  &  \textbf{3.160}  & \textbf{0.244}   & \textbf{0.930}  & \textbf{0.249}   & \textbf{0.935}  &  \textbf{0.594}     & \textbf{0.364}  \\
    \bottomrule
\end{tabular}
\caption{\textbf{Quantitative results for objective evaluation. }Comparison to five established baselines and the groundtruth with nine quantitative metrics. Here, Au. stands for the audio output capability, and Vd. indicates supporting variable-duration music.}
\label{table1}
\end{table*}

\subsection{Transition-Beat Aligner and Adapter} 
To achieve precise rhythmic consistency where visual scene transitions coincide with music beats, we first introduce the transition-beat aligner. As shown in Fig. \ref{fig:fig2}, frame-level video parsing provides scene transitions, denoted by the binary indicator $f^{frame-v_i}_{b}$, where a value of 1 signifies a transition and 0 indicates its absence. Concurrently, we apply an RNN-based beat detector \cite{bock2016madmom} to generate a corresponding binary sequence $f^{frame-m_i}_{b}$, indicating frame-wise music beats. Both sequences operate at a consistent frame rate of 16 fps. The intersection $f^{frame_i}_{b}=f^{frame-v_i}_{b}\cap f^{frame-m_i}_{b}$ identifies the timestamps where visual transitions align with music beats, thereby ensuring cross-modal rhythmic consistency. To extract the aligned frame-level rhythmic features highlighted by the intersected sequence $\hat f^{frame_i}_{b}$ from visual inputs, a ResNet(2+1)D-18 model \cite{du2018r2ds} is trained using binary cross-entropy (BCE) loss over $N$ number of samples: 
\begin{equation}
\begin{split}
    \mathcal{L}_{BCE}= 
    -\dfrac{1}{N}\sum_{i=1}^N[f^{frame_i}_{b}log(\hat f^{frame_i}_{b})\\
    +(1-f^{frame_i}_{b})log(1-\hat f^{frame_i}_{b})] 
\end{split}
\label{eq8}
\end{equation}
After training, the transition-beat aligner is capable of predicting a timestamp mask indicating the presence or absence of transition-beat matches in the target music. We extract activations from the penultimate layer and interpolate them to align with the temporal resolution of the music latent $z$, which is subsequently processed by the transition-beat adapter. Although concatenation along the channel dimension is feasible, it risks overemphasizing conditional signals, potentially distorting the music latent $z$ of the generative model. Drawing inspiration from adaptive normalization layers (AdaLNs) \cite{xu2019understanding}, we propose the transition-beat adapter to ensure precise alignment between generated music and designated rhythmic features. Specifically, we normalize the music feature $z_i$ into a scale $\gamma_i$ and a shift $\beta_i$ based on AdaLNs with the two zero-initialized convolution layers, where $\gamma_i$ and $\beta_i$ are learned from the transition-beat aligner. The adaptive normalization layers are integrated into each encoder block of the U-TNet architecture, with $\gamma_i$ and $\beta_i$ modulating $z_i$ by a linear projection:
\begin{equation}
    z_i = z_i + \gamma_i \cdot z_i + \beta_i 
\end{equation}

\subsection{Train and Inference}

In the training phase, we first pre-train the music reconstruction VAE model and the transition-beat aligner independently (dashed lines in Fig. \ref{fig:fig2}). We then freeze these components, along with the frozen text and video encoders. Subsequently, the full latent diffusion is trained with only the trainable time embedder, facilitating the model to focus on semantic and temporal details from hierarchical video representation. The transition-beat module is excluded in this stage to prioritize conditioned music generation. Finally, we integrate the pre-trained aligner into the framework and jointly optimize the adapter to refine rhythmic consistency. The training configurations are provided in Appendix C.1. 


During inference, latent music diffusion receives random noise as the initial $z_T$. Hierarchical video parsing processes input videos to provide conditional information represented by the encoders for the generative latent diffusion. The transition-beat aligner predicts visual features correlated with transition-beat events, which are incorporated into the music latent via the adapter (Fig. \ref{fig:fig2}, solid lines).

\begin{table*}[ht]
\centering
\begin{tabular}{l@{\hspace{35pt}}rrrrrrr}
    \toprule 
            & \multicolumn{2}{c}{Preference Rate} & \hspace{10pt} & \multicolumn{4}{c}{Preference Score}                            \\
            \cmidrule(lr){2-3} \cmidrule(lr){5-8}
            & \multicolumn{2}{c}{Top-1} &    & \multicolumn{2}{c}{MOS-Q} 
            & \multicolumn{2}{c}{MOS-A}      \\
            & Expert  & Non-expert &   & Expert  & Non-expert & Expert  & Non-expert \\
    \midrule
CMT\cite{di2021video}          & 3.625\%      & 2.000\%   &
                    & 5.622\textsubscript{$\pm$0.213}
                    & 6.139\textsubscript{$\pm$0.329}
                    & 4.680\textsubscript{$\pm$0.247}    
                    & 4.924\textsubscript{$\pm$0.189}         \\
Diff-BGM\cite{li2024diff}  & 2.250\%      & 2.125\%   &
                    & 5.406\textsubscript{$\pm$0.185}
                    & 5.935\textsubscript{$\pm$0.314}
                    & 4.387\textsubscript{$\pm$0.243}
                    & 4.530\textsubscript{$\pm$0.212}         \\
M$^2$UGen\cite{liu2023m}        & 5.375\%         & 5.125\%   &
                    & 5.340\textsubscript{$\pm$0.162}
                    & 5.863\textsubscript{$\pm$0.307}
                    & 5.814\textsubscript{$\pm$0.221}    
                    & 6.127\textsubscript{$\pm$0.205}         \\
VidMuse\cite{tian2024vidmuse}     & 4.250\%      & 2.750\%  &
                    & 4.767\textsubscript{$\pm$0.234}
                    & 4.992\textsubscript{$\pm$0.128}
                    & 5.467\textsubscript{$\pm$0.229}    
                    & 5.270\textsubscript{$\pm$0.210}         \\
GVMGen\cite{zuo2025gvmgen}     & 11.125\%      & 10.125\%  &
                    & 5.418\textsubscript{$\pm$0.223}
                    & 5.693\textsubscript{$\pm$0.262}
                    & 6.467\textsubscript{$\pm$0.197}    
                    & 6.374\textsubscript{$\pm$0.251}         \\
Ours                & \textbf{73.375\%}     & \textbf{77.875\%}     &  
                    & \textbf{6.892}\textsubscript{$\pm$0.173}
                    & \textbf{7.537}\textsubscript{$\pm$0.195}     
                    & \textbf{7.341}\textsubscript{$\pm$0.174}    
                    & \textbf{7.852}\textsubscript{$\pm$0.260}         \\
    \bottomrule
\end{tabular}
\caption{\textbf{Qualitative results for subjective evaluation. }The preference rates in the Top-1 rank and the preference scores in MOS-Q and MOS-A with CI95 for expert and non-expert groups.}
\label{table2}
\end{table*}

\section{Experiments}
\subsection{Dataset and Settings}

\textbf{Dataset.} We introduce TB-Match, a high-quality video-music paired dataset comprising around 18,000 samples sourced from e-commerce advertisements and video-sharing platforms. 
This type of video typically exhibits frequent and highly precise synchronization between scene transitions and music beats, rendering them especially suitable for studying temporal and rhythmic alignment in video-music relationships. Each pair undergoes rigorous hybrid filtering, combining automated quality control (e.g., minimum SNR of 20dB, visual-auditory rhythmic coherence, and emotional consistency) with manual expert curation to ensure strong video-music relevance. The details of the dataset can be found in Appendix B. Furthermore, we incorporate the M$^2$UGen \cite{liu2023m} dataset, contributing 13,000 video-music pairs, resulting in approximately 280 hours of total training data. For evaluation, we reserve a validation set of 1,000 TB-Match samples, ensuring no overlap with training data. For the universality study, we supplement the SymMV dataset \cite{zhuo2023video}, Sora-generated silent videos \cite{brooks2024video}, and other random data.


\textbf{Implementation.}
We leverage a pre-trained VAE and vocoder \cite{liu2024audioldm}, fine-tuning for our specific task. Modality-specific encoders, excluding the timing embedder, are frozen during the entire training process. T-UNet architecture adheres to the configuration described in \cite{liu2024audioldm}, employing a 1000-step diffusion process. To handle variable-length inputs, we standardize music clips to durations between 10-60 seconds. Audio signals are downsampled to 16 kHz and transformed into Mel-spectrograms using 60 frequency bins with a hop size of 256. The video inputs are processed at 16 fps.

\textbf{Baseline Models.}
We conduct a comparative evaluation with five state-of-the-art methods: GVMGen \cite{zuo2025gvmgen}, VidMuse \cite{tian2024vidmuse}, M$^2$UGen \cite{liu2023m}, Diff-BGM \cite{li2024diff} and CMT \cite{di2021video}. GVMGen employs hierarchical attentions to align spatial-temporal video-music features. VidMuse adopts long-short-term modeling to capture the temporal dependencies. M$^2$UGen leverages LLMs to handle cross-modal relationships. Diff-BGM addresses semantic and temporal alignment at the clip level, and the CMT adapts rhythmic features to the generated music. The output of M$^2$UGen is restricted to approximately 10 seconds. For fair comparison, we loop the shorter segments to match the duration of the videos. Diff-BGM and CMT produce variable-length MIDI representations, which we convert to waveform audio via high-quality synthesizers to ensure format consistency across all evaluated methods.

\subsection{Objective Evaluation}
\textbf{Metrics.} This section outlines the quantitative metrics employed to evaluate the generated music on four dimensions: musical quality, semantic alignment, temporal synchronization, and rhythmic consistency.

\textbf{Music Quality.} 
We adopt three metrics to evaluate fidelity in generation tasks \cite{agostinelli2023musiclm}. Inception Score (IS) measures the diversity and the perceptual clarity of generated spectrograms compared to the groundtruth. Fréchet Audio Distance (FAD) quantifies the distance between the embedding distributions of generated and reference samples. Kullback-Leibler Divergence (KLD) assesses similarity by comparing probability distributions derived from activations of a pre-trained Musicnn model \cite{pons2019musicnn}.

\textbf{Semantic Alignment.}
The Contrastive Language-Audio Pretraining (CLAP) score \cite{wu2023largescale} quantifies the semantic alignment between audio signals and corresponding textual descriptions. To directly assess visual-audio consistency, we employ the pre-trained LanguageBind model \cite{zhu2023languagebind}, which projects video and music into a unified textual latent space. The cosine distance between embeddings is calculated to produce the LanguageBind (LB) score.

\textbf{Temporal Synchronization.} The video-music semantics remain consistent over time for temporal synchronization. Since VeM explicitly captures temporal dynamics through storyboard sequences, we compute time-weighted CLAP and LB scores (tw-CLAP and tw-LB). The weight of each storyboard $i$ is proportional to its relative duration ($d_{i}$/$d_{total}$, \emph{storyboard duration / total duration}).

\begin{table}
    \centering
    \begin{tabular}{lc@{\hspace{5pt}}c@{\hspace{5pt}}c@{\hspace{5pt}}c@{\hspace{5pt}}c@{\hspace{5pt}}c}
 \toprule 
        & \multicolumn{2}{c}{SymMV} & \multicolumn{2}{c}{Sora} & \multicolumn{2}{c}{Others} \\
        \cmidrule(lr){2-3} \cmidrule(r){4-5} \cmidrule(r){6-7}
        & LB$\uparrow$  & TB$_{IoU}$$\uparrow$ & LB$\uparrow$  & TB$_{IoU}$$\uparrow$ & LB$\uparrow$  & TB$_{IoU}$$\uparrow$ \\
\midrule
CMT       & 0.912      & 0.314    & 0.758      & 0.671    & 0.578      & 0.337      \\
Diff      & 0.643      & 0.253    & 0.898      & 0.667   & 0.589      & 0.325   \\
M$^2$U    & 0.925      & 0.296    & 1.029      & 0.725   & 0.885      & 0.332    \\
Vid       & 0.787      & 0.312    & 0.982     & 0.785   & 0.670      & 0.400    \\
GVM       & 0.910      & 0.260   & 1.084      & 0.814   & 0.887      & 0.391       \\
Ours      & \textbf{0.989}     & \textbf{0.331}     & \textbf{1.106}     & \textbf{0.829}    & \textbf{0.895}     & \textbf{0.453}         \\
\bottomrule
\end{tabular}
\caption{\textbf{Universality evaluation on other data with three quantitative metrics.} Diff, M$^2$U, Vid and GVM stand for Diff-BGM, M$^2$UGen, VidMuse and GVMGen, respectively.}
\label{table3}
\end{table}

\textbf{Rhythmic Consistency.} Rhythmic consistency requires that video transitions align with music beats. Assuming that the ideal video-music pairs are well-synchronized, we introduce the Beats Intersection over Union (IoU) metric, B$_{IoU}$. It measures the overlap, within a specified threshold, between the number of detected beats in generated music $B_{syn}$ and that in the groundtruth $B_{gt}$, defined as:
\begin{equation}
    B_{IoU}=\dfrac{B_{gt}\cap B_{syn}}{B_{gt}\cup B_{syn}}
\end{equation}
Furthermore, we present the Transitions-Beats IoU metric, TB$_{IoU}$, which calculates the intersection within a threshold between the video transition timestamps $T_{v}$ and the music beat timestamps $B_{m}$. The temporal threshold in both the beat and the transition detectors is 0.5 seconds, and the detectors are detailed in Section 3.2. The score is defined as:
\begin{equation}
    TB_{IoU}=\dfrac{T_{v}\cap B_{m}}{T_{v}\cup B_{m}}
\end{equation}

\begin{table*}[ht]
    \centering
    \begin{tabular}{cccrcccc@{\hspace{8pt}}c@{\hspace{8pt}}ccc}
    \toprule
        HVP-Cond & SG-CAtt & TB-As & IS $\uparrow$ & FAD$\downarrow$  & KLD$\downarrow$ & CLAP$\uparrow$  & LB$\uparrow$  & tw-CLAP$\uparrow$  & tw-LB$\uparrow$  & B$_{IoU}$$\uparrow$ & TB$_{IoU}$$\uparrow$      \\
    \midrule
        $\times$ & $\times$ & $\times$  & 0.823  & 6.692  & 4.714  & 0.180  & 0.624   & 0.188  & 0.625    & 0.221  &0.197\\
        $\times$ & $\times$ & $\checkmark$   & 0.772  & 7.217 & 5.097 & 0.172 
        & 0.639  & 0.181  & 0.643  & 0.433  &0.283\\
        $\checkmark$ & $\checkmark$ & $\times$ & 1.191 & 4.382 & 3.608 &0.231 
        & 0.890  & 0.236  & 0.882  & 0.403  &0.265\\
        $\checkmark$ & $\times$ & $\times$ & 1.140 & 5.712  & 3.869  & 0.218  & 0.735   & 0.227  & 0.742   & 0.383   & 0.220\\
        $\checkmark$ & $\checkmark$ & $\checkmark$  & \textbf{1.263}  & \textbf{4.043}  &  \textbf{3.160}  & \textbf{0.244}   & \textbf{0.930}  & \textbf{0.249}   & \textbf{0.935}  &  \textbf{0.594}     & \textbf{0.364} \\
    \bottomrule
    \end{tabular}
    \caption{ \textbf{Ablation study on three components} including hierarchical video parsing conditions (HVP-Cond), storyboard-guided cross-attention (SG-CAtt), and transition-beat aligner and adapter (TB-As). Nine quantitative metrics are employed. }
    \label{table4}
\end{table*}

\textbf{Quantitative Results.}
Table \ref{table1} presents the comparative evaluation with five baselines in nine quantitative metrics, where the audio output (Au.) and the variable duration (Vd.) are emphasized. Our approach consistently outperforms existing methods, showcasing improvements in music quality, semantic alignment, temporal synchronization, and rhythmic consistency. VeM surpasses not only audio-based (GVMGen, VidMuse, and M$^2$UGen) but also MIDI-based methods (CMT and Diff-BGM), which is particularly notable for two reasons: 1) the inherent decoupling of MIDI allows the integration of fine-grained musical details during generation, and 2) the generated MIDI is converted to audio via high-quality synthesizers, effectively reducing auditory noise. Meanwhile, the time-weighted CLAP and LB scores exceed their non-weighted counterparts in our approach, demonstrating the local semantic and temporal alignment within storyboards. Overall, the proposed method exhibits superior quality, enhancing the audio-visual experience.
 
\textbf{Universality Study.}
To assess universality, we conduct experiments in external domains, distinct from our training set. As shown in Table \ref{table3}, VeM outperforms baselines across diverse inputs, indicating its effectiveness even in zero-shot scenarios. Partial results are presented due to space constraints, and complete results are provided in Appendix D.1.

\subsection{Subjective Evaluation}

Due to the subjective nature of video-music alignment evaluation, we conduct a human study with 50 participants, divided into expert and non-expert groups. The expert group consists of 5 film production experts and 25 professional musicians. The non-experts include 20 amateur viewers. 16 video samples are involved, each with 6 variations featuring soundtracks generated by different methods. Participants watch the 6 versions in a randomized order. The preference rate is reported as the probability that a soundtrack receives the top rank (Top-1). Meanwhile, participants evaluate each video on two 10-point Likert scales (1 = worst, 10 = best) to assess music quality and video-music alignment. Results are reported as mean-opinion-scores for quality (MOS-Q) and alignment (MOS-A), along with 95\% confidence intervals (CI95). The details are provided in Appendix D.2.

\textbf{Qualitative Results.} Table \ref{table2} presents the comprehensive subjective evaluation, demonstrating the consistent superiority of the proposed method. Specifically, VeM achieves the highest Top-1 preference rate among both expert and non-expert participants. For mean opinion scores, MOS-Q and MOS-A scores indicate superior perceived music quality and video-music alignment. The performance advantages across evaluator backgrounds underscore the effectiveness.

\subsection{Ablation Study}
We conduct ablation studies to analyze the contribution of each component within the proposed framework. The components include hierarchical video parsing conditions (HVP-Cond), storyboard-guided cross-attention (SG-CAtt), transition-beat aligner and adapter (TB-As). Table \ref{table4} details five ablated variants. The unconditional generation removes all conditional signals. To assess the impact of TB-As, we exclude both HVP-Cond and SG-CAtt. 
We further evaluate the combined influence of HVP-Cond and SG-CAtt by omitting the fine-grained rhythmic synchronization from TB-As. The effectiveness of SG-CAtt is tested by substituting it with standard cross-attention. Lastly, we present the results for the complete VeM model that incorporates all components.

\begin{figure}[ht]
\centering
\includegraphics[width=\linewidth]{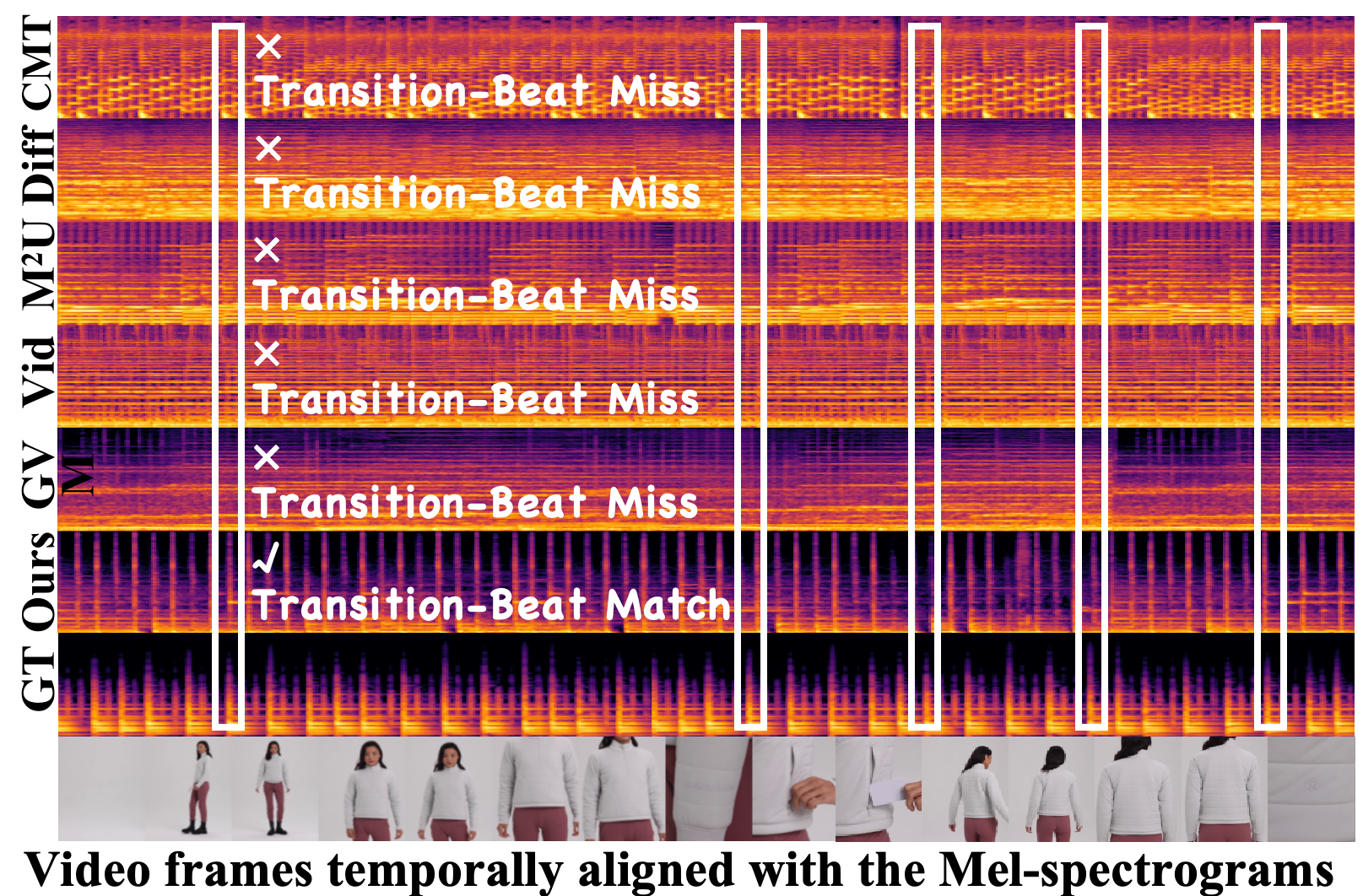}
\caption{\textbf{Visualized comparison} shows Mel-spectrograms alongside the video frames from different methods.} 
\label{fig:fig3}
\end{figure}

The variant utilizing only TB-As (w/TB-As) achieves the highest transition-beat alignment measured by B$_{IoU}$ and TB$_{IoU}$, highlighting the importance of the TB-As module for fine-grained rhythmic synchronization. Compared to the variant with only TB-As, the one incorporating both HVP-Cond and SG-CAtt (w/HVP-Cond \& SG-CAtt) outperforms on the rest metrics, indicating the substantial contribution of HVP-Cond and SG-CAtt to the semantic and temporal alignment. Replacing SG-CAtt with standard cross-attention (w/HVP-Cond) results in degenerate performance, confirming the superiority of the SG-CAtt mechanism. The complete VeM demonstrates the best overall performance, validating the cumulative contribution of each component.

\subsection{Visualization of Generated Music}
Fig. \ref{fig:fig3} visualizes Mel-spectrograms of audio samples alongside the video frames. Compared with baselines, VeM exhibits greater consistency with the groundtruth spectrogram, particularly in preserving temporal and rhythmic dynamics corresponding to salient visual scene transitions, highlighted by the white bounding boxes in Fig. \ref{fig:fig3}.

\section{Conclusion}

In this paper, we propose VeM, a latent music diffusion to generate high-quality soundtracks semantically, temporally, and rhythmically aligned with video. VeM leverages hierarchical video parsing to comprehensively capture rich details for generation. Storyboard-guided cross-attention facilitates semantic alignment and temporal synchronization. Fine-grained rhythmic precision is achieved by the transition-beat aligner and adapter. Experimental results on a constructed video-music dataset with novel evaluation metrics showcase superior performance. Future work will explore video-integrated music editing and investigate more sophisticated alignment techniques.

\section{Acknowledgments}
This research was supported in part by the National Key Laboratory of General Artificial Intelligence (Beijing Institute for General Artificial Intelligence), the PKU-Alimama Joint Research Laboratory, Special Program of National Natural Science Foundation of China (Grant No. T2341003), Advanced Discipline Construction Project of Beijing Universities, and Major Program of National Social Science Fund of China (Grant No. 21ZD19).


\setcounter{secnumdepth}{2} 
\appendix

\section{Details of Hierarchical Video Parsing}
\label{appendix_sec_a}

Comprehensive video analysis requires the extraction of five key elements:

1) Overarching theme, atmosphere, and emotional impact

2) Smooth video segmentation into coherent storyboards

3) Narrative and visual compositions within storyboards

4) Temporal boundaries and duration of storyboards

5) Precise timing of frame-level visual changes

\begin{figure}[h]
\centering
\includegraphics[width=\linewidth]{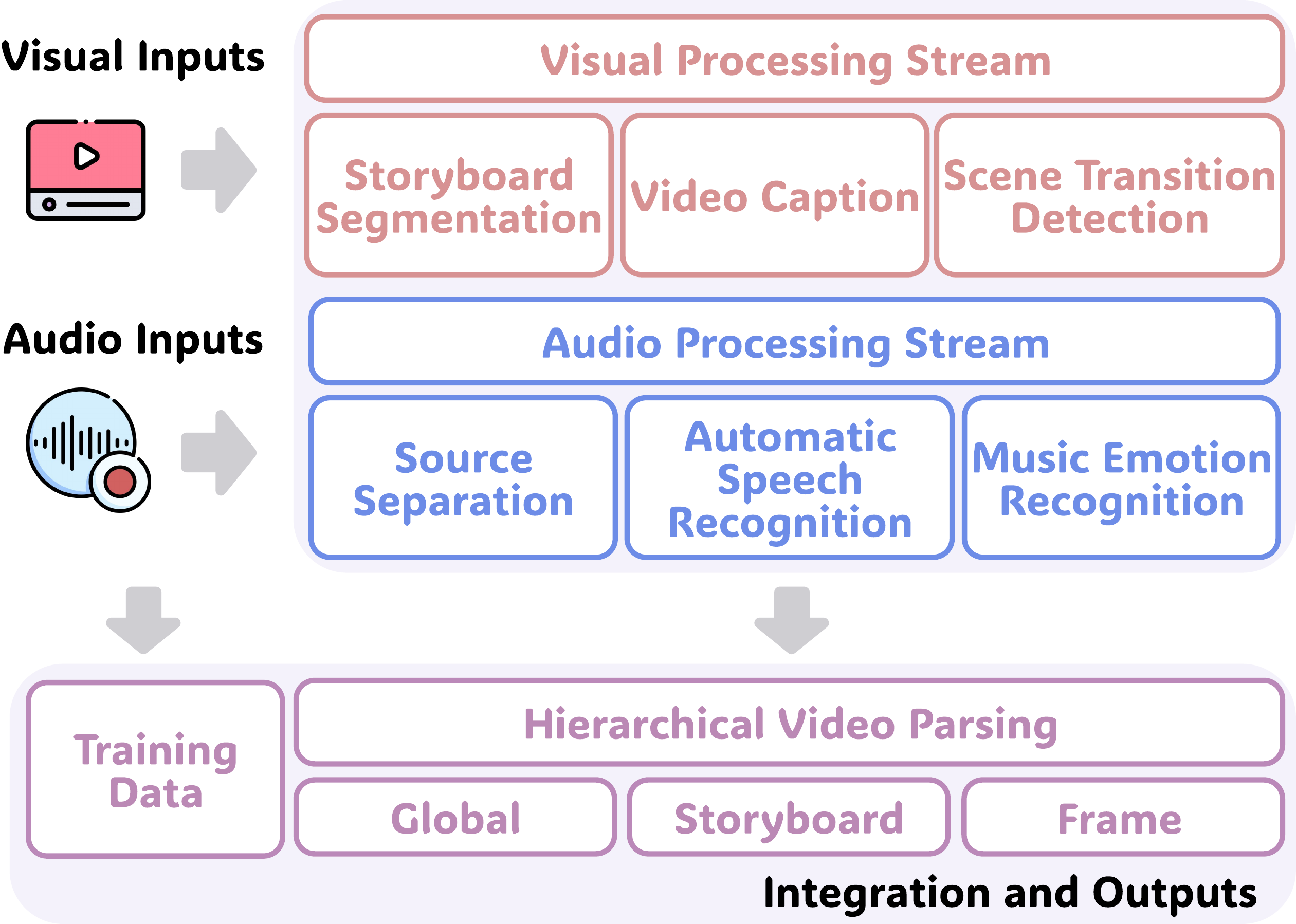}
\caption{Implementation of hierarchical video parsing.} 
\label{appendix_fig:fig1}
\end{figure}

As introduced in Section 3.2, these five elements are collectively derived from hierarchical video parsing. This section outlines the implementation details. As illustrated in Fig. \ref{appendix_fig:fig1}, our automated pipeline processes video and corresponding audio through a dual-path processing stream to extract complementary, temporally aligned multimodal representations.

\noindent\textbf{Visual Processing Stream.} The visual stream deconstructs the video into temporal structure and semantic content, structured as follows:

\begin{itemize}
    \item \textbf{Video Storyboard Segmentation}: We employ a ResNet(2+1)D-18 model \cite{du2018r2ds} for storyboard boundary detection, computing frame-wise dissimilarity to partition the video into coherent storyboards with precise start and end timestamps. This step addresses Key 2 (smooth segmentation) and Key 4 (temporal boundaries), forming the foundation for cross-modal temporal alignment.
    \item \textbf{Scene Transition Detection}: PySceneDetect \cite{Castellano_PySceneDetect} automatically identifies frame-level transitions within each storyboard, refining the temporal structure. The combined output, storyboard boundaries, and transition timestamps construct the complete scene transition timelines for Key 5 (frame-level timing).
    \item \textbf{Video Caption}: The entire video and individual storyboards are processed by Qwen-VL-7B \cite{Qwen2.5-VL}, a large vision-language model, to generate global and local semantic descriptions. The captions link visual content to textual narrative, addressing Key 1 (theme/emotion) and Key 3 (narrative compositions).
\end{itemize}

\noindent\textbf{Audio Processing Stream.} The audio stream analyzes the audio track to extract narrative and musical characteristics, involving three components:

\begin{itemize}
    \item \textbf{Source Separation}: For mixed speech-music tracks, we employ a source separation method, Spleeter \cite{spleeter2020}, to separate the original audio into a clean vocal and an instrumental music track.
    \item \textbf{Automatic Speech Recognition (ASR)}: The vocal track is transcribed by FunASR \cite{gao2023funasr}, producing timestamped textual transcription synchronized with visual content, which enhances Key 3 (narrative compositions).
    \item \textbf{Music Emotion Recognition}: The separated instrumental track is processed by a music understanding model \cite{yu2023musicagent}. We train the model to annotate music with 50 emotional/stylistic tags (e.g., uplifting, tense, romantic, acoustic), characterizing Key 1 (theme/emotion).
\end{itemize}

\noindent\textbf{Integration and Output.} The hierarchical parsing integrates outputs from both streams and aligns timestamps at three levels: 
\begin{itemize}
    \item \textbf{Global}: Video captions (Qwen-VL-7B) and music tags (Music Emotion Recognition) resolve Key 1.
    \item \textbf{Storyboard}: Storyboard boundaries (ResNet(2+1)D), storyboard descriptions (Qwen-VL-7B) with corresponding video frames, and ASR transcriptions address Keys 2–4.
    \item \textbf{Frame}: Scene transitions (PySceneDetect) ensure Key 5.

\end{itemize}
This hierarchical multimodal representation serves as structured and machine-readable training data for the generative model in the proposed method.  

\section{Dataset Details}
\label{appendix_sec_b}
This section details the constructed video-music paired dataset, TB-Match, sourced from two domains: e-commerce advertisements and public video-sharing platforms that are specifically chosen because they generally exhibit frequent scene transitions and cuts. The adjacent scenes often show significant changes while preserving a coherent thematic consistency throughout the video. It imposes higher demands on the precise alignment between video and music in semantic, temporal, and rhythmic dimensions.

\begin{figure}[t]
\centering
\includegraphics[width=\linewidth]{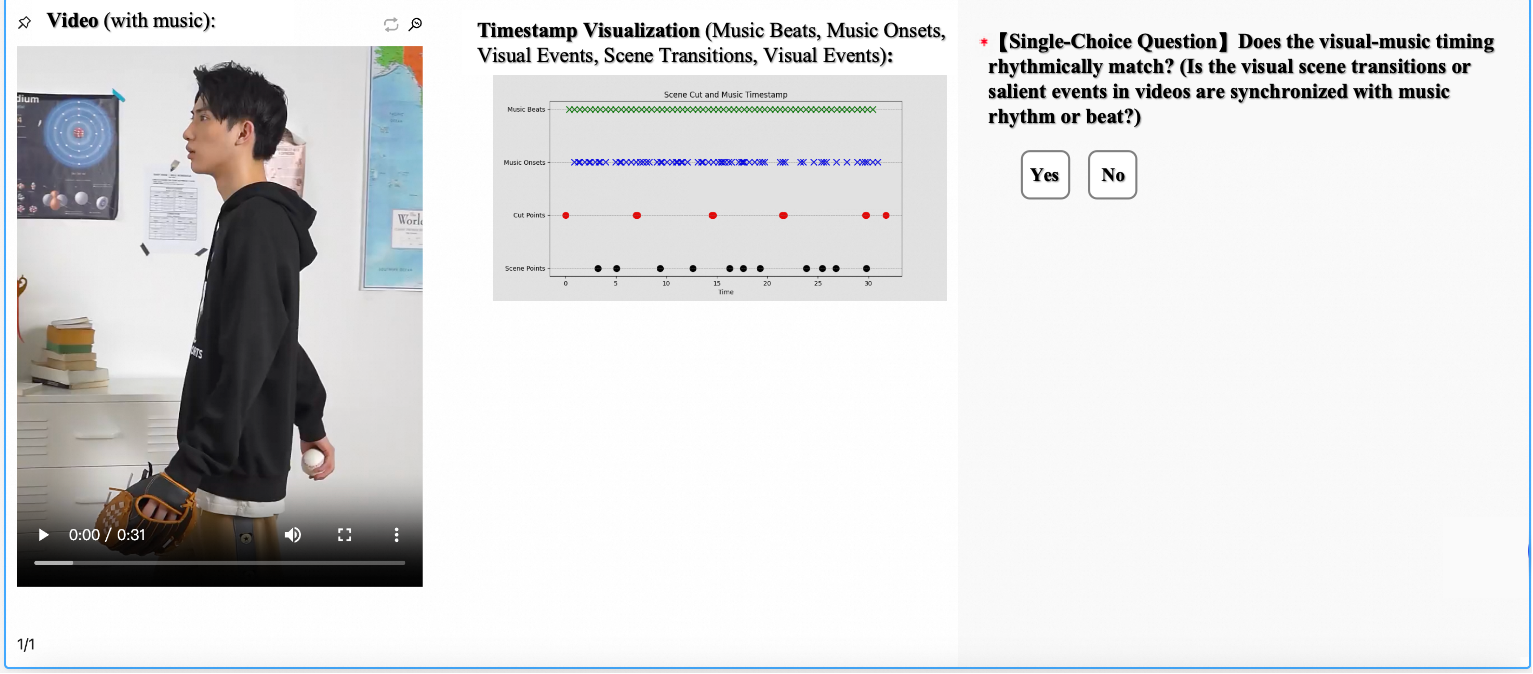}
\caption{Screenshot of manual annotation tool.} 
\label{appendix_fig:fig2}
\end{figure}

\textbf{E-commerce Subset.}
We first collect over 30,000 e-commerce advertisement videos with corresponding background music, covering diverse domains including apparel/jewelry, food/grocery, furniture/appliances, sports/outdoors products, etc. An initial filtering process removed videos containing voiceover, with a minimum Signal-to-Noise Ratio (SNR) below 20 dB, exceeding 120 seconds in duration, or containing more than 20 shots. This yields approximately 16,000 videos ranging from 10 to 120 seconds in length. To further ensure strong alignment between visual transitions and music beats, we adopt a dual-path approach:

\noindent1) \textbf{Manual Annotation}: 20 music conservatory students are invited in subjective evaluations on the 16,000 videos. As depicted in Fig. \ref{appendix_fig:fig2}, annotators watch the original videos with music and visualization to mark the corresponding timestamps for music beat onsets, video transition points, and salient visual events. They are guided to assess whether visual transitions and salient events are synchronized with the music beat. This process results in the selection of around 8,000 high-quality videos with reliable visual-music alignment.

\noindent2) \textbf{Algorithmic Alignment}: For the remaining videos, we identify beat timestamps and scene transition timestamps \cite{bock2016madmom}. Transitions are algorithmically aligned to the nearest beats. Temporal interpolation between frames is applied to ensure precise synchronization of transitions with the beat.

\textbf{Video-sharing-platform Subset.}
We collect an additional 800 high-quality, royalty-free advertisement videos from YouTube featuring cityscapes, natural scenery, sports activities, brand showcases, etc. To maintain the duration distribution of the e-commerce subset, longer videos are segmented into 20-60 second clips based on music beat timestamps, ensuring each clip contains 2-20 shots. This yields nearly 2,000 additional samples.

Ultimately, TB-Match comprises approximately 18,000 high-quality video-music pairs, with video durations ranging from 10 to 120 seconds and no video exceeding 20 shots. To facilitate downstream tasks, we provide hierarchical video parsing annotations corresponding to each video. Additionally, we integrate the M$^2$UGen, contributing an extra 13,000 pairs. Our complete training dataset encompasses approximately 280 hours.

\section{Implementation Details}

\subsection{Configuration}
\label{appendix_sec_c_1}
We train the music VAE using the Adam optimizer, with a learning rate of $10^{-6}$ and a batch size of 32 for a minimum of 1.5 million steps on a single NVIDIA V100 GPU. Besides the music from our collected TB-Match dataset, we also utilize music datasets such as MusicCaps and Million Song Dataset. The transition-beat aligner is trained on the collected TB-Match dataset, which has strong music-visual relevance, for over 0.8 million steps across 8 NVIDIA V100 GPUs with a batch size of 64. The full VeM model is trained for 2.8 million steps on a single NVIDIA A100 GPU, with a batch size of 4 on the TB-Match dataset. The learning rate is $10^{-5}$. It should be noted that we limit our batch size because of the scarcity of GPUs, potentially restricting the model performance.

\subsection{Influence of Video Encoders}
To identify a more effective video encoder, we experiment with two pre-trained models: MAViL \cite{NEURIPS2023mavil} and LanguageBind \cite{zhu2023languagebind}. Both methodologies align video and audio in latent space, yet differ in approach: MAViL aligns video and audio directly within a shared latent representation, whereas LanguageBind aligns both modalities to a semantically consistent textual space.

Table \ref{appendix_table1} presents a comparative analysis revealing distinct performance characteristics. MAViL excels in music quality metrics, whereas LanguageBind has a slight advantage in semantic alignment evaluation. Crucially, MAViL shows significantly better performance on rhythmic alignment metrics, indicating its superior ability to maintain the temporal structure of video and ensure generated music adheres to rhythmic patterns. Based on these outcomes, we choose MAViL as our video encoder. Simultaneously, we introduce the pre-trained LanguageBind model as an evaluation metric to assess semantic relevance between video content and generated music in our results.

\begin{table}[h]
    \centering
    \begin{tabular}{cc@{\hspace{5pt}}c@{\hspace{5pt}}c@{\hspace{5pt}}c@{\hspace{5pt}}c@{\hspace{5pt}}c}
 \toprule 
        
        & IS$\uparrow$  & FAD$\downarrow$ & KLD$\downarrow$  & CLAP$\uparrow$ & B$_{IoU}$$\uparrow$  & TB$_{IoU}$$\uparrow$ \\
\midrule
MA.       & \textbf{1.257}      & \textbf{4.229}    & 3.231      & \textbf{0.230}    &\textbf{ 0.484}      & \textbf{0.353}      \\
LB.       & 1.255      & 4.278    & \textbf{3.199}      & 0.217    & 0.473      & 0.267   \\
\bottomrule
\end{tabular}
\caption{Comparison of two different video encoders with six quantitative metrics.}
\label{appendix_table1}
\end{table}

\begin{table*}[th]
    \centering
    \begin{tabular}{l|lccccccccc}
 \toprule 
       Datasets  & Methods  & IS $\uparrow$ & FAD$\downarrow$  & KLD$\downarrow$ & CLAP$\uparrow$  & LB$\uparrow$  & tw-CLAP$\uparrow$  & tw-LB$\uparrow$  & B$_{IoU}$$\uparrow$ & TB$_{IoU}$$\uparrow$  \\
 \midrule
 \multirow{6}{*}{SymMV}
&CMT       & 1.211           & 5.658           & 3.771 
& 0.235    & 0.912           & 0.216           & 0.914      
& 0.425    & 0.314 \\

&Diff-BGM  & 1.169           & 6.632           & 5.247   
& 0.152    & 0.643           & 0.169           & 0.703     
&  0.327   & 0.253 \\

&M$^2$UGen & 1.184            & 5.832           & 4.813  
& 0.218    & 0.925           & 0.223            & 0.917     
& 0.327    & 0.296 \\

&VidMuse   & 1.176           & 7.197           & 5.010  
&  0.194   & 0.787           &  0.203          & 0.790     
& 0.398    & 0.312 \\

&GVMGen    & 1.193           & 6.334           & 3.943  
&  0.232   & 0.910           &  0.249          & 0.922     
& 0.410    & 0.260 \\
&Ours      & \textbf{1.224}  & \textbf{5.547}        & \textbf{3.753}                 
& \textbf{0.268}  & \textbf{0.989}        & \textbf{0.279}                 
& \textbf{0.991}  &  \textbf{0.475}       & \textbf{0.331}  \\
\hline\hline
\multirow{6}{*}{Sora}
&CMT       & -           & -           & - 
& 0.163    & 0.758           & 0.158           & 0.772      
& -   & 0.671 \\

&Diff-BGM   & -           & -           & - 
& 0.171    & 0.898           & 0.178           & 0.906     
& -   & 0.667 \\

&M$^2$UGen  & -           & -           & - 
& 0.243    & 1.029           & 0.251            & 1.083     
& -    & 0.725 \\

&VidMuse    & -           & -           & - 
& 0.186    & 0.982           &  0.192          & 0.979     
& -    & 0.785 \\

&GVMGen     & -           & -           & - 
&  0.271   & 1.084           &  0.270          & 1.106     
& -    & 0.814 \\
&Ours       & -           & -           & - 
& \textbf{0.286}  & \textbf{1.106}        & \textbf{0.292}                 & \textbf{1.117}  &  -       & \textbf{0.829}  \\
\hline\hline
\multirow{6}{*}{Others} 
&CMT       & 1.167           & 7.151           & 5.476 
& 0.117    & 0.578           & 0.119           & 0.574      
& 0.423    & 0.337 \\

&Diff-BGM  & 1.168           & 6.840           & 5.326   
& 0.121    & 0.589           & 0.124           & 0.592     
& 0.416    & 0.325 \\

&M$^2$UGen & 1.209            & 6.272           & 4.015  
& 0.185    & 0.822           & 0.182            & 0.831     
& 0.407    & 0.332 \\

&VidMuse   & 1.178           & 7.267           & 4.891  
&  0.134   & 0.670           &  0.128          & 0.668     
& 0.584    & 0.400 \\

&GVMGen    & 1.212           & 5.879           & 3.948  
&  0.218   & 0.887           &  0.223          & 0.873     
& 0.572    & 0.391 \\
&Ours      & \textbf{1.247}  & \textbf{5.532}        & \textbf{3.792}                 
& \textbf{0.237}  & \textbf{0.895}        & \textbf{0.241}                 
& \textbf{0.901}  &  \textbf{0.625}       & \textbf{0.453}  \\
\bottomrule
\end{tabular}
\caption{Universality evaluation on SymMV, Sora videos, and other video data with quantitative and qualitative metrics.}
\label{appendix_table2}
\end{table*}

\subsection{Influence of Inference Steps}
\begin{figure}[th]
\centering
\includegraphics[width=\linewidth]{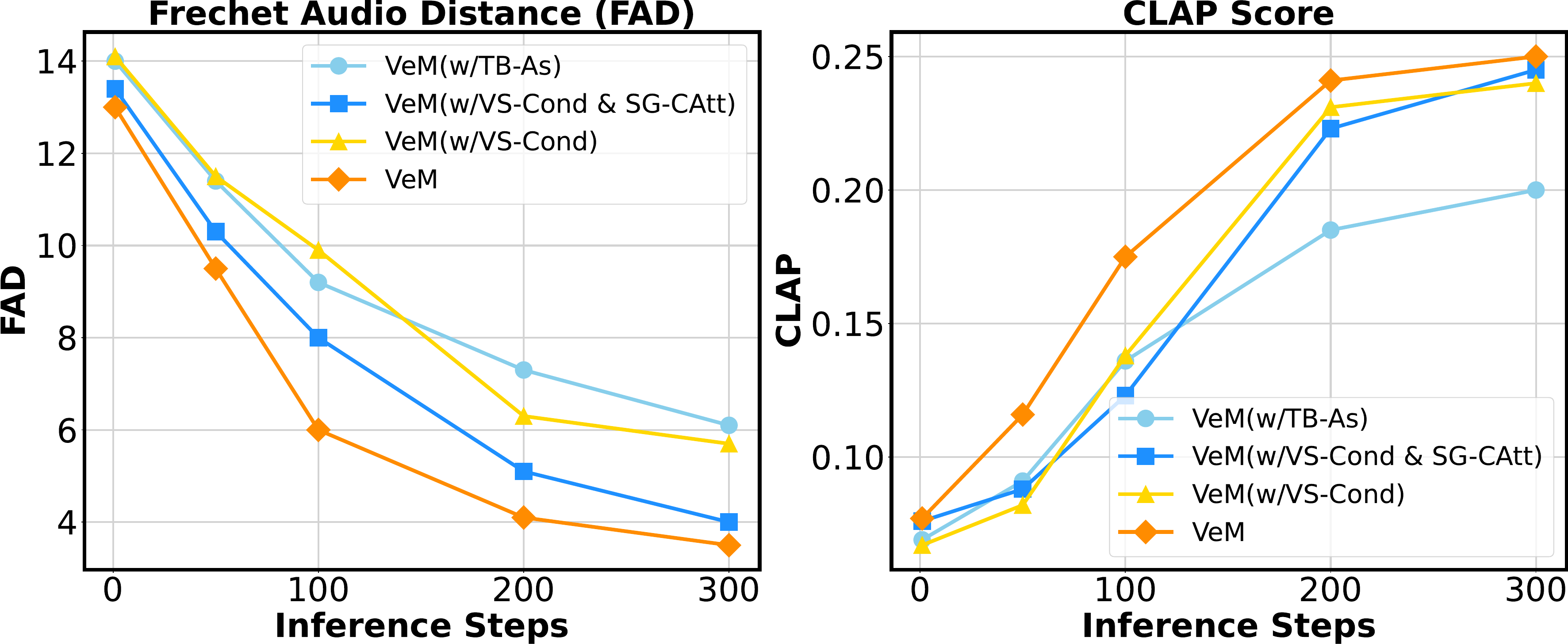}
\caption{Different inference steps of ablated variants.} 
\label{appendix_fig:fig3}
\end{figure}

We further investigate the performance under reduced sampling steps from 300 to 50 and a single step. Fig. \ref{appendix_fig:fig3} illustrates the observed trends of four ablated variants based on the FAD and LB metrics across various inference sampling steps. Performance degradation is observed for VeM(w/ TB-As), VeM(w/ HVP-Cond \& SG-CAtt), VeM(w/ HVP-Cond), and VeM. Specifically, the reduced sampling steps lead to a noticeable decrease in performance from 200 to 50, with a further reduction to a single step resulting in severely diminished scores. All music is generated using 200 steps.

\section{Evaluation}
\subsection{Universality Evaluation}
\label{appendix_sec_d_1}
To assess universality, we conduct experiments on SymMV \cite{zhuo2023video}, Sora-generated silent videos \cite{brooks2024video}), and other video data. Due to space constraints in the main manuscript, we only report results on the LB and TB$_{IoU}$ metrics. Table \ref{appendix_table2} demonstrates the comprehensive universality study across nine objective and two subjective metrics in three distinct testing regimes. Note that several metrics could not be computed for Sora-generated videos due to the absence of groundtruth. 

\begin{figure*}[ht]
\centering
\includegraphics[width=\linewidth]{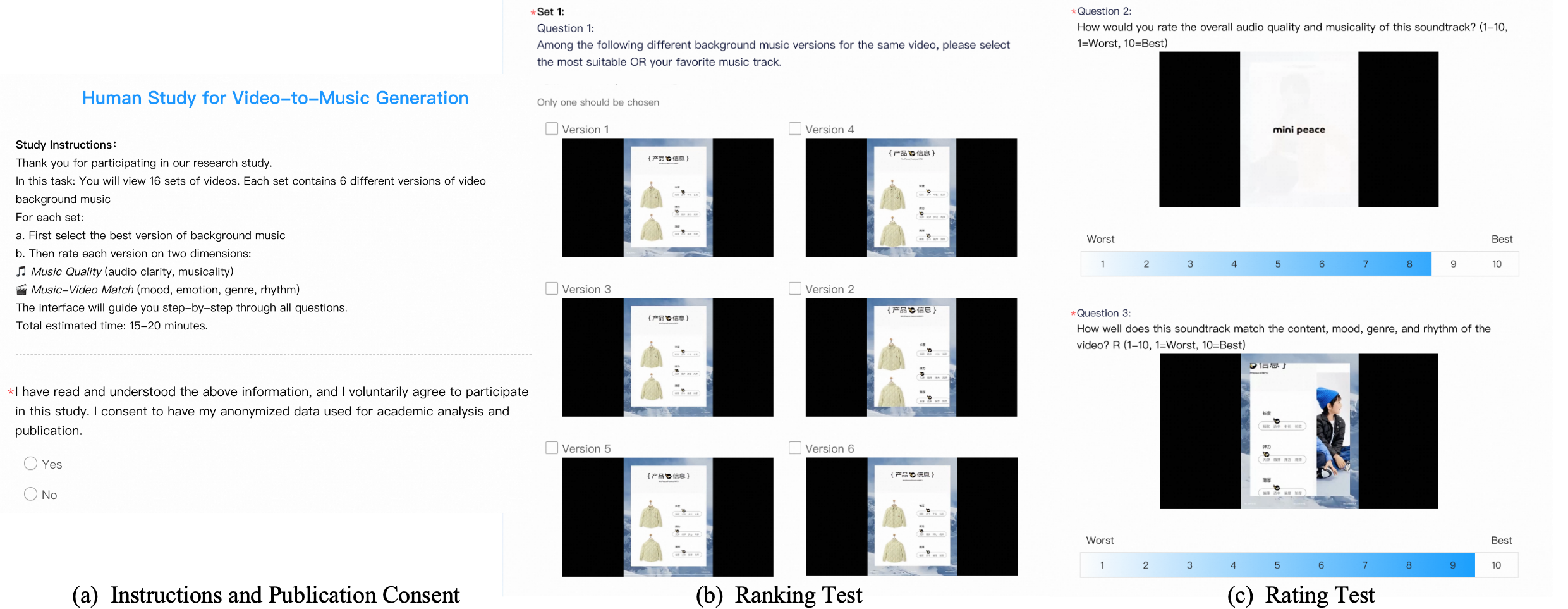}
\caption{Screenshots of human study interface.} 
\label{appendix_fig:fig4}
\end{figure*}

\subsection{Subjective Evaluation}
\label{appendix_sec_d_2}
We conduct a human study with 50 participants, explicitly divided into two groups for subjective evaluation:

\begin{itemize}
    \item \textbf{Expert Group} (30 participants) comprises 5 film professionals with over five years of experience in roles such as film editing, sound design, and directing; and 25 professional musicians with over two years of experience, including composers, performers, and music producers.

    \item \textbf{Non-Expert Group} (20 participants) consists of amateur viewers, aged 18-45, with diverse backgrounds and no specialized training in film or music production. They report regular consumption of videos and various music tastes, recruited to represent the general audiences.
\end{itemize}

Participants are guided to evaluate 16 distinct test video samples in a group. Each sample presents 6 variations: soundtracks generated by the proposed VeM and five comparative methods. The evaluation process utilizes a user-friendly interface shown in Fig. \ref{appendix_fig:fig4} for straightforward interaction. For each test video, participants perform two tasks in sequence:

\begin{itemize}
    \item \textbf{Ranking}: Participants rank the 6 versions in randomized order according to overall preference. The preference rate is reported as the probability that a soundtrack receives the top rank (Top-1, except the groundtruth).

    \item \textbf{Rating}: Participants independently score each version on two 10-point Likert scales (1 = worst, 10 = best). The scales assess two dimensions: 1) Music Quality: How would you rate the overall audio quality and musicality of this soundtrack? 2) Video-Music Alignment: How well does this soundtrack match the content, mood, genre, and rhythm of the video? The results are reported as Mean Opinion Scores for Music Quality (MOS-Q) and Mean Opinion Scores for Video-Music Alignment (MOS-A), along with their corresponding 95\% confidence intervals.
\end{itemize}

\subsection{Demo Description} 

\begin{links}
  \link{Demos}{https://vem-paper.github.io/VeM-page/}
\end{links}
A subset of generated samples is available on the anonymous link, and the highly compressed version is provided in the \textbf{Supplementary Material}, comprising three components:

\begin{itemize}
    \item \textbf{Comparison on TB-Match Video Test}: We provide 5 video samples, each with 7 soundtrack variations generated by 5 baseline methods (CMT, Diff-BGM, M2UGen, VidMuse, GVMGen) and 2 for the proposed VeM. 
    To showcase the superior temporal and rhythmic alignment of our method, we introduce a specialized demo (VeM\_click) for each test video, totaling 35 presented videos. The audible clicks clearly mark music beat timestamps that coincide with video transitions (the intervals between two timestamps fall below the 0.5s threshold). These markers show that transitions occur exactly at beat boundaries rather than at intermediate beat phases. This explicitly verifies that VeM achieves tighter beat-transition synchronization compared to baselines.

    \item \textbf{Cross-Domain Video Test}: We present 13 test video demos from various external domains, including 3 SymMV samples and 10 randomly selected online ones. The audible click markers emphasize beat-transition alignment. These demos demonstrate robustness in complex and diverse scenarios, particularly in terms of semantic relevance and rhythmic consistency.
    
    \item \textbf{Sora-Generated Video Test}: We display 10 demos (2 variations per video) for 5 silent videos generated by Sora. Prioritizing temporal continuity in Sora-generated videos results in fewer transitions, and thus, click markers are omitted for this set. Each video is tested twice with distinct music outputs, both tracks maintaining consistent style, ambiance, feelings, and temporal structure.

\end{itemize}

\section{Limitations}

Since music beats typically remain constant over extended temporal segments, achieving perfect alignment of every visual transition with a music beat is inherently challenging. Consequently, frequent transitions inevitably lead to partial alignment failures. Our future work intends to address this limitation by concentrating on multimodal change points, specifically examining concurrent visual and auditory shifts at transition boundaries. Furthermore, we consider that music generation methods should prioritize edit-ability for practical deployment. Our future research will consequently focus on developing enhanced editing capabilities while maintaining alignment ability.

\bibliography{aaai2026}

\end{document}